\documentclass[superscriptaddress,showpacs,aps,twocolumn]{revtex4}
\usepackage{amssymb}
\usepackage{amsmath}
\usepackage{graphicx}
\usepackage{color}
\usepackage{amsfonts}
\usepackage{multirow}
\usepackage{tabularx}

\begin{document}

\title{ Surface analysis via fast atom diffraction: pattern visibility and spot-beam
contribution}
\author{L. Frisco}
\affiliation{Dpto. de F\'{\i}sica, FCEN, Universidad de Buenos Aires, Buenos Aires,
Argentina}
\author{J. E. Miraglia}
\affiliation{Instituto de Astronom\'{\i}a y F\'{\i}sica del Espacio (IAFE, CONICET-UBA),
casilla de correo 67, sucursal 28, C1428EGA, Buenos Aires, Argentina}
\affiliation{Dpto. de F\'{\i}sica, FCEN, Universidad de Buenos Aires, Buenos Aires,
Argentina}
\author{M. S. Gravielle}
\affiliation{Instituto de Astronom\'{\i}a y F\'{\i}sica del Espacio (IAFE, CONICET-UBA),
casilla de correo 67, sucursal 28, C1428EGA, Buenos Aires, Argentina}

\begin{abstract}
Grazing incidence fast atom diffraction (GIFAD or FAD) is a
sensitive tool for surface analysis, which strongly relies on the
quantum coherence of the incident beam. In this article the
influence of the incidence conditions and the projectile mass on the
visibility of the FAD patterns is addressed. Both parameters
determine the transverse coherence length of the impinging
particles, which governs the general features of FAD distributions.
We show that by varying the impact energy, while keeping the same
collimating setup and normal energy, it is possible to control the
interference mechanism that prevails in FAD patterns. Furthermore,
we demonstrate that the contribution coming from different positions
of the focus point of the incident particles, which gives rise to
the spot-beam effect, allows projectiles to explore different zones
of a single crystallographic channel when a narrow surface area is
coherently lighted.  In this case the spot-beam effect gives also
rise to a non-coherent background, which contributes to the gradual
quantum-classical transition of FAD spectra. Present results are
compared with available experimental data, making evident that the
inclusion of focusing effects is necessary for the proper
theoretical description of the experimental distributions.
\end{abstract}

\date{\today }

\maketitle

\section{Introduction}

Over the years surface analysis techniques involving collisions with atomic
particles have strongly contributed to the characterization of the surface
properties of solids. Among them, grazing-incidence fast atom diffraction
(GIFAD or FAD), developed in the last decade \cite{Schuller2007,Rousseau2007}%
, can be considered as one of the most sensitive methods to investigate the
morphological and electronic characteristics of ordered surfaces \cite%
{Winter2011}. FAD is a versatile analysis technique that can be applied to a
wide variety of materials \cite%
{Debiossac2014,Busch2009,Rios2013,Winter2009,Schuller2009b,Seifert2013,Zugarramurdi2015,Momeni2018}%
, providing structural parameters of the topmost atomic layer with an
extraordinary accuracy \cite%
{Schuller2010,Schuller2012,Seifert2013sc,Seifert2016,DelCueto2017,Debiossac2017}.

Since the use of FAD as a surface analysis tool requires of both the
observation of well-resolved interference structures and its
appropriate theoretical description, an essential aspect is the
degree of quantum coherence of the incident beam, which governs the
general shape of FAD patterns. The degree of coherence of the
incident particles depends on the collimating setup and the
incidence conditions .
In Refs. \cite%
{Seifert2015,Gravielle2015,Gravielle2016} it was shown that for a
given collision system, with fixed incidence energy and angle, the
experimental collimating scheme determines the overall features of
the projectile distribution, allowing one to examine two different
interference mechanisms - inter-channel or intra-channel
interferences - by varying the size of the collimating slit. This
behavior is related to the transverse length of the surface area
that is coherently lighted by the incident beam, whose knowledge
becomes crucial for an appropriate comparison between experiments
and simulations.

In an equivalent way, the incidence conditions are expected to
affect the interference spectra produced via FAD by using a given
collimating setup \cite{Moix2011,Minniti2012}. In this article we
explore the influence of the energy and mass of the impinging
projectile \cite{Gravielle2017}, as well as the width of the
incidence channel, on the visibility of FAD patterns obtained with a
fixed collimating aperture. Furthermore, the contribution of the
spot-beam effect, associated with random-distributed focus points of
the incident particles, is addressed. We demonstrate that this
effect allows projectiles to probe different regions of the
atom-surface potential when the transverse coherence length of the
impinging atoms is smaller than the width of the channel. But in
this case the spot-beam effect gives rise to a non-coherent
background, which strongly modifies the visibility of the
interference structures, contributing to the transition from quantum
to classical projectile distributions.

The study is confined to fast He and Ne atoms grazingly impinging on
LiF(001) along the $\left\langle 110\right\rangle $ and $\left\langle
100\right\rangle $ channels. In order to derive the extent of the surface
region that is coherently illuminated by the atomic beam after collimation
we resort to the Van Cittert-Zernike theorem \cite{BornWolf, Gravielle2016}.
This information is then used to determine the size of the coherent initial
wave packet to be evolved within the Surface-Initial Value Representation
(SIVR) approximation \cite{Gravielle2014}. The SIVR approach is a
semi-quantum method that has proved to provide a successful description of
experimental FAD patterns for different collision systems \cite%
{Gravielle2015,Bocan2016,Bocan2018}, offering a clear account of the
different interference mechanisms. In this version of the SIVR approximation
we also include the variation of the relative position of the focus point
(wave-packet center) of the incident particles on the crystal surface, which
gives rise to the spot-beam effect.

The paper is organized as follows: The theoretical formalism, including the
spot-beam contribution, is summarized in Sec. II. Results for different
incidence conditions - incidence energy and channel - and projectile masses
are presented and discussed in Secs. III.A and III.B, respectively. In Sec.
III.C the contribution of the spot-beam effect is analyzed, while in Sec.
III.D we study the gradual quantum-classical transition of the projectile
distributions. Finally, Sec. III.E our results are contrasted with available
experimental data and in Sec. IV we outline our conclusions. Atomic units
(a.u.) are used unless otherwise stated.

\section{Theoretical model}

In this work we extend the previous SIVR model \cite{Gravielle2014} to deal
with different focus points of the incident particles. The relative position
of the focus point of the beam, with respect to the surface lattice sites,
plays a negligible role when the transverse coherence length of the
impinging particles is longer than or equal\ to the width of the incidence
channel. But it gains importance as the transverse coherence length
decreases. Since it is not experimentally possible to control the focus
position of the incident projectiles at such an accuracy level, we consider
that each atomic projectile impacts on the surface plane at a different
position $\mathbf{R}_{s}$, which coincides with the central position of the
initial coherent wave packet.

\begin{figure}[tbp]
\includegraphics[width=0.5\textwidth]{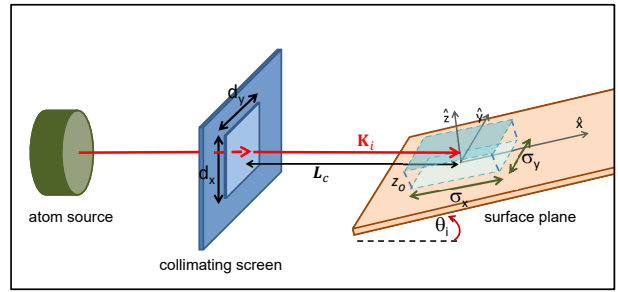}
\caption{Depiction of the collimating scheme, together with the reference
frame.}
\label{esquema}
\end{figure}

For a given position $\mathbf{R}_{s}$ of the focus point, the SIVR
scattering amplitude for the elastic transition $\mathbf{K}_{i}\rightarrow
\mathbf{K}_{f}$, $\mathbf{K}_{i}$ ($\mathbf{K}_{f}$) being the initial
(final) momentum of the atomic projectile, with $\left\vert \mathbf{K}%
_{f}\right\vert =\left\vert \mathbf{K}_{i}\right\vert $, can be expressed as
\cite{Gravielle2014}
\begin{eqnarray}
A_{if}^{{\small (SIVR)}}(\mathbf{R}_{s}) &=&\int d\mathbf{r}_{o}\ f_{i}(%
\mathbf{r}_{o}-\mathbf{R}_{s})  \notag \\
&&\times \int d\mathbf{k}_{o}\ g_{i}(\mathbf{k}_{o})\ a_{if}^{{\small (SIVR)}%
}(\mathbf{r}_{o},\mathbf{k}_{o}),  \label{aif}
\end{eqnarray}%
where $a_{if}^{{\small (SIVR)}}(\mathbf{r}_{o},\mathbf{k}_{o})$ is the
partial transition amplitude, given by Eq. (9) of Ref. \cite{Gravielle2014},
which is associated with the classical projectile path $\boldsymbol{r}%
_{t}\equiv \boldsymbol{r}_{t}(\mathbf{r}_{o},\mathbf{k}_{o})$, with $\mathbf{%
r}_{o}$ and $\mathbf{k}_{o}$ being the starting position and momentum,
respectively, at the time $t=0$. In Eq. (\ref{aif}) functions $f_{i}(\mathbf{%
r}_{o}-\mathbf{R}_{s})$ and $g_{i}(\mathbf{k}_{o})$ describe the spatial and
momentum profiles of the initial coherent wave packet at a fixed distance $%
z_{o}$ from the surface where the atomic projectile is hardly affected by
the surface interaction. The frame of reference is located on the first
atomic layer, with the $\widehat{x}$ versor along the incidence channel and
the $\hat{z}$ versor oriented perpendicular to the surface, aiming towards
the vacuum region (see Fig. \ref{esquema}). Within this reference frame, the
starting position at $t=0$ can be expressed as $\mathbf{r}_{o}=\mathbf{r}%
_{o}^{\prime }+z_{o}\widehat{z}$, where $\mathbf{r}_{o}^{\prime }=$ $x_{o}%
\widehat{x}+y_{o}\widehat{y}$ is the component parallel to the surface plane
and the normal distance $z_{o}$ is chosen as equal to the lattice constant.

To derive the spatial profile of the initial wave packet we assume that the
atomic beam is produced by an extended incoherent quasi-monochromatic
source, placed at a long distance from a rectangular collimating aperture,
with sides $d_{x}$ and $d_{y}$. The collimating slit is oriented
perpendicular to the momentum $\mathbf{K}_{i}$, \ in such a way that the
side of length $d_{y}$ is parallel to the surface (i.e. parallel to the $%
\widehat{y}$ versor), while the side of length $d_{x}$ forms an angle $%
\theta _{x}=\pi /2-\theta _{i}$ with the surface plane (i.e. with the $%
\widehat{x}$ versor), with $\theta _{i}$ being the glancing incidence angle,
as depicted in Fig. \ref{esquema}. \ The function $\ f_{i}(\mathbf{r}%
_{o}^{\prime }-\mathbf{R}_{s})$ is here obtained from the complex degree of
coherence by applying the Van Cittert-Zernike theorem \cite{BornWolf}, as
explained in Refs. \cite{Gravielle2015,Gravielle2016}. Under the condition
of extended source, given by Eqs. (A.9) and (A.10) of Ref. \cite%
{Gravielle2016}, it \ can be approximate by means of normalized Gaussian
functions $G\left[ \omega ,x\right] =[2/(\pi \omega ^{2})]^{1/4}\exp
(-x^{2}/\omega ^{2})$, as
\begin{equation}
f_{i}(\mathbf{r}_{o}^{\prime }-\mathbf{R}_{s})\simeq G\left[ \sigma
_{x},x_{o}-X_{s}\right] G\left[ \sigma _{y},y_{o}-Y_{s}\right] ,
\label{profile-x}
\end{equation}%
where
\begin{equation}
\sigma _{x}=\frac{L_{c}\lambda _{\bot }}{\sqrt{2}d_{x}},\qquad \sigma _{y}=%
\frac{L_{c}\lambda }{\sqrt{2}d_{y}},  \label{sigmax}
\end{equation}%
denote the \textit{transverse} \textit{coherence} \textit{lengths} of the
initial coherent wave packet \cite{Tonomura1986} along the $\widehat{x}$- $\
$and $\widehat{y}$- directions, respectively, and the two-dimensional vector
$\mathbf{R}_{s}=$ $X_{s}\widehat{x}+Y_{s}\widehat{y}$ corresponds to the
central position of the wave packet. In Eq. (\ref{sigmax}), $L_{c}$ \ is the
collimator-surface distance, $\lambda =2\pi /K_{i}$ is the de Broglie
wavelength of the impinging atom, \ and $\lambda _{\bot }=\lambda /\sin
\theta _{i}$ is the perpendicular wavelength associated with the initial
motion normal to the surface plane. The momentum profile $g_{i}(\mathbf{k}%
_{o})$ is derived from Eq. (\ref{profile-x}) by applying the Heisenberg
uncertainty relation, reading \cite{Gravielle2015,Gravielle2016}
\begin{equation}
g_{i}(\mathbf{k}_{o})\simeq g_{i}(\Omega _{o})=G(\omega _{\theta },\theta
_{o}-\theta _{i})G(\omega _{\varphi },\varphi _{o}),\text{ \ \ \ }
\label{profile-angle}
\end{equation}%
where $\Omega _{o}\equiv (\theta _{o},\varphi _{o})$ is the solid angle
associated with the $\mathbf{k}_{o}$- direction, $k_{0}=K_{i}$, and%
\begin{equation}
\omega _{\theta }=\frac{d_{x}}{\sqrt{2}L_{c}}\text{, \ }\omega _{\varphi }=%
\frac{d_{y}}{\sqrt{2}L_{c}}.  \label{wtitafi}
\end{equation}

Taking into account that FAD patterns are essentially produced by the
interference of a single projectile with itself, contributions to the
scattering probability coming from different focus points of the impinging
particles\ must be added incoherently. Thence, the differential scattering
probability in the direction of the solid angle $\Omega _{f}$ can be
obtained from Eq. (\ref{aif}), except for a normalization factor, as
\begin{equation}
\frac{dP^{{\small (SIVR)}}}{d\Omega _{f}}=\int d\mathbf{R}_{s}\left\vert
A_{if}^{{\small (SIVR)}}(\mathbf{R}_{s})\right\vert ^{2},  \label{proba}
\end{equation}%
where $\Omega _{f}\equiv (\theta _{f},\varphi _{f})$ is the solid angle
corresponding to the $\mathbf{K}_{f}$- direction, with $\theta _{f}$ the
final polar angle, measured with respect to the surface, and $\varphi _{f}$
the azimuthal angle, measured with respect to the $\widehat{x}$ axis. In Eq.
(\ref{proba}), the $\mathbf{R}_{s}$- integral involves different relative
positions within the crystal lattice, covering an area equal to a reduced
unit cell of the surface.

\section{Results}

To study the effect of the coherence length on FAD patterns, in Ref. \cite%
{Gravielle2015} the size of the collimating aperture was varied by
maintaining a fixed incidence condition, i.e., 1 keV He atoms impinging on a
LiF(001) surface along $\left\langle 110\right\rangle $ with $\theta
_{i}=0.99$ deg, in order to compare with the available experiments \cite%
{Seifert2015}. The goal of this work is to extend\ such research by
analyzing the influence of the impact energy, the incidence channel and the
projectile mass \cite{Gravielle2017}, as well as the contribution of the
spot-beam effect, for a given collimating setup. For this purpose we examine
final angular distributions of $^{4}$He and $^{20}$Ne atoms elastically
scattered from LiF(001) along the $\left\langle 110\right\rangle $ \ and $%
\left\langle 100\right\rangle $ channels, after passing through a square
collimating aperture with $d_{x}=d_{y}=0.2$ mm, situated at a distance $%
L_{c}=25$ cm from the surface plane \cite{Seifert2015}. For both projectiles
the surface-atom interaction was evaluated with an improved pairwise
additive potential \cite{Miraglia2017}, which includes non-local terms of
the electronic density in the kinetic, exchange and correlation energies.
The potential model also takes into account projectile polarization and
rumpling effects. In turn, for the numerical evaluation of the SIVR
transition probability we employed the MonteCarlo technique to solve the
six-dimensional integral involved in Eqs. (\ref{aif}) and (\ref{proba}),
i.e., on $\mathbf{r}_{o}^{\prime }\equiv (x_{o},y_{o})$, $\Omega _{o}\equiv
(\theta _{o},\varphi _{o})$, and $\mathbf{R}_{s}\equiv (X_{s},Y_{s})$, using
about $10^{7}$ points. Each of these points involves a further time
integration along the classical path, included in $a_{if}^{{\small (SIVR)}}(%
\mathbf{r}_{o},\mathbf{k}_{o})$, which was evaluated with a step-adaptive
integration method \cite{Gravielle2014}.

\subsection{Influence of the incidence conditions}

We start studying the dependence of the general features of the FAD patterns
on the total energy $E$, with $E=K_{i}^{2}/(2m_{P})$ and $m_{P}$ being the
projectile mass. Due to the fast velocity of the projectile along the
incidence channel, which makes its parallel motion\ mainly sensitive to the
average potential in this direction, FAD patterns from LiF surfaces are
essentially governed by the normal energy $E_{\bot }=E\sin ^{2}\theta _{i}$%
,\ which is\ associated with the slow motion of the atom in the
perpendicular plane \cite{Winter2011}. Along this article, except in Sec.
III. E, we have kept the normal energy $E_{\bot }=0.3$ eV as a constant for
the different impact energies.

\begin{figure*}[tbp]
\includegraphics[width=0.8 \textwidth]{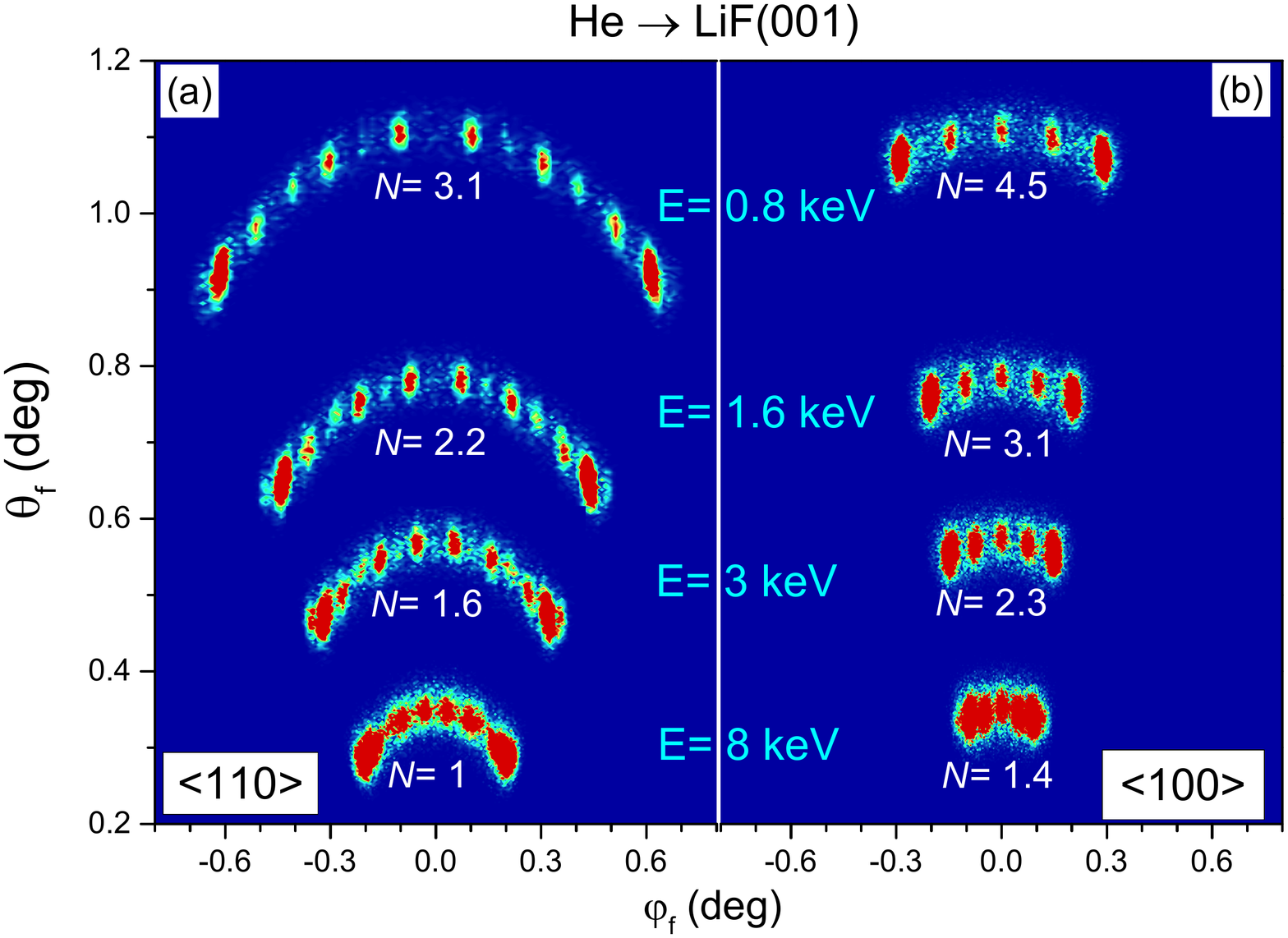}
\caption{(Color online) Two-dimensional projectile distributions, as a
function of $\protect\theta _{f}$ and $\protect\varphi _{f}$, for He atoms
impinging on LiF(001) along the (a) $\left\langle 110\right\rangle $ and (b)
$\left\langle 100\right\rangle $ directions. The helium beam is collimated
by means of a square aperture of sides $d_{x}=d_{y}=$ \ $0.2$ mm and the
normal energy is $E_{\bot }=0.3$ eV. In both panels, angular distributions
for different impact energies - $E=$ \ $0.8$, $1.6$, $3$, and $8$ keV - are
shown, indicating the corresponding $N$ values, as given by Eq. (\protect\ref%
{ny}).}
\label{hemap}
\end{figure*}

In Fig. \ref{hemap} we show $dP^{{\small (SIVR)}}/d\Omega _{f}$, as a
function of $\theta _{f}$ and $\varphi _{f}$, for He projectiles scattered
along the (a) $\left\langle 110\right\rangle $ \ and (b) $\left\langle
100\right\rangle $ directions with different impact energies, ranging from $%
0.8$ to $8$ keV. Since neither inelastic processes nor the detector
resolution function were taken into account in the present SIVR
calculations, for a given channel and $E_{\perp }$- value, all the angular
distributions are expected to display the same number of interference
maxima, independently of the total impact energy \cite{Schuller2009}.
However, in Fig. \ref{hemap} this behavior is verified for incidence along $%
\left\langle 100\right\rangle $ only, while on the contrary, the
distributions corresponding to the $\left\langle 110\right\rangle $
direction display interference peaks whose number and relative intensities
depend \ strongly on $E$. This unexpected fact can be understood in terms of
the number of equivalent parallel channels that are coherently illuminated
by the atomic beam, as it \ will be discussed below.

It is now well-established that the structures of FAD spectra come from the
combination of inter-\ and intra\textit{-} channel interferences \cite%
{Winter2011}. Each of these mechanisms is associated with a different factor
of the SIVR transition amplitude \cite{Gravielle2014}: The inter\textit{-}%
channel factor, produced by interference among equivalent trajectories
running along different parallel channels, which gives rise to equally
spaced and intense Bragg peaks, and the intra\textit{-}channel factor, due
to interference inside a single channel, which originates supernumerary
rainbow maxima \cite{Schuller2008,Winter2011}. Accordingly, when the surface
area coherently lighted by the atomic beam covers a region containing an
array of parallel channels, FAD patterns display Bragg peaks whose
intensities are modulated by the intra-channel factor. But when only one
channel is coherently illuminated, the spectra present supernumerary rainbow
peaks, without any trace of Bragg interference. Thence, the number $N$ of \
coherently lighted channels results a critical parameter that determines the
general shape of FAD distributions. It can be roughly estimated from the
transverse coherence length $\ $of the incident particles as
\begin{equation}
N\simeq \frac{2\sigma _{y}}{a_{y}}=\frac{\sqrt{2}L_{c}}{d_{y}}\frac{2\pi }{%
a_{y}K_{i}}\;,  \label{ny}
\end{equation}%
where $\sigma _{y}$ is given by Eq. (\ref{sigmax}) and $a_{y}$ denotes the
width of the incidence channel, with $a_{y}=5.4$ a.u. ($a_{y}=3.8$ a.u.) for
$\left\langle 110\right\rangle $ ($\left\langle 100\right\rangle $).

For a given collimating setup, the value of $N$ varies with both the impact
energy, through its dependence on $K_{i}$, and the incidence direction,
through the channel width, as given by Eq. \ (\ref{ny}). In Fig. \ref{hemap}
(a), corresponding to the $\left\langle 110\right\rangle $ direction,\ the
application of Eq. (\ref{ny}) for the lowest energy - $E=0.8$ keV - leads to
$N=3.1$ parallel channels coherently lighted by the He beam, which gives
rise to\ a projectile distribution with well separated Bragg peaks \cite%
{Schuller2009}. But when $E$ augments, and consequently, $N$ decreases,
these Bragg maxima broaden \cite{Gravielle2014}, causing the interference
structures for $E=1.6$ keV to become comparatively wider than those for $%
E=0.8$ keV. In Fig. \ref{hemap} (a) the Bragg peaks for $\left\langle
110\right\rangle $ \ incidence start to blur out for a total energy about $3$
keV, for which $N=1.6$, while the limit case corresponding to \textit{pure}
intra-channel interference is reached at $E=8$ keV. At this energy a single $%
\left\langle 110\right\rangle $ channel is coherently illuminated by the
incident beam, producing a projectile distribution with rainbow and
supernumerary rainbow maxima only. In contrast with this strong dependence
on $E$ of the $\left\langle 110\right\rangle $ patterns, in Fig. \ref{hemap}
(b), for the same impact energies but along $\left\langle 100\right\rangle $
all the spectra display a constant number of Bragg peaks (i.e., 5 peaks), in
accord with $N$ values higher than $1$, varying from $N=4.5$ to $1.4$ for
the lowest and highest energies, respectively.

\begin{figure}[tbp]
\includegraphics[width=0.4\textwidth]{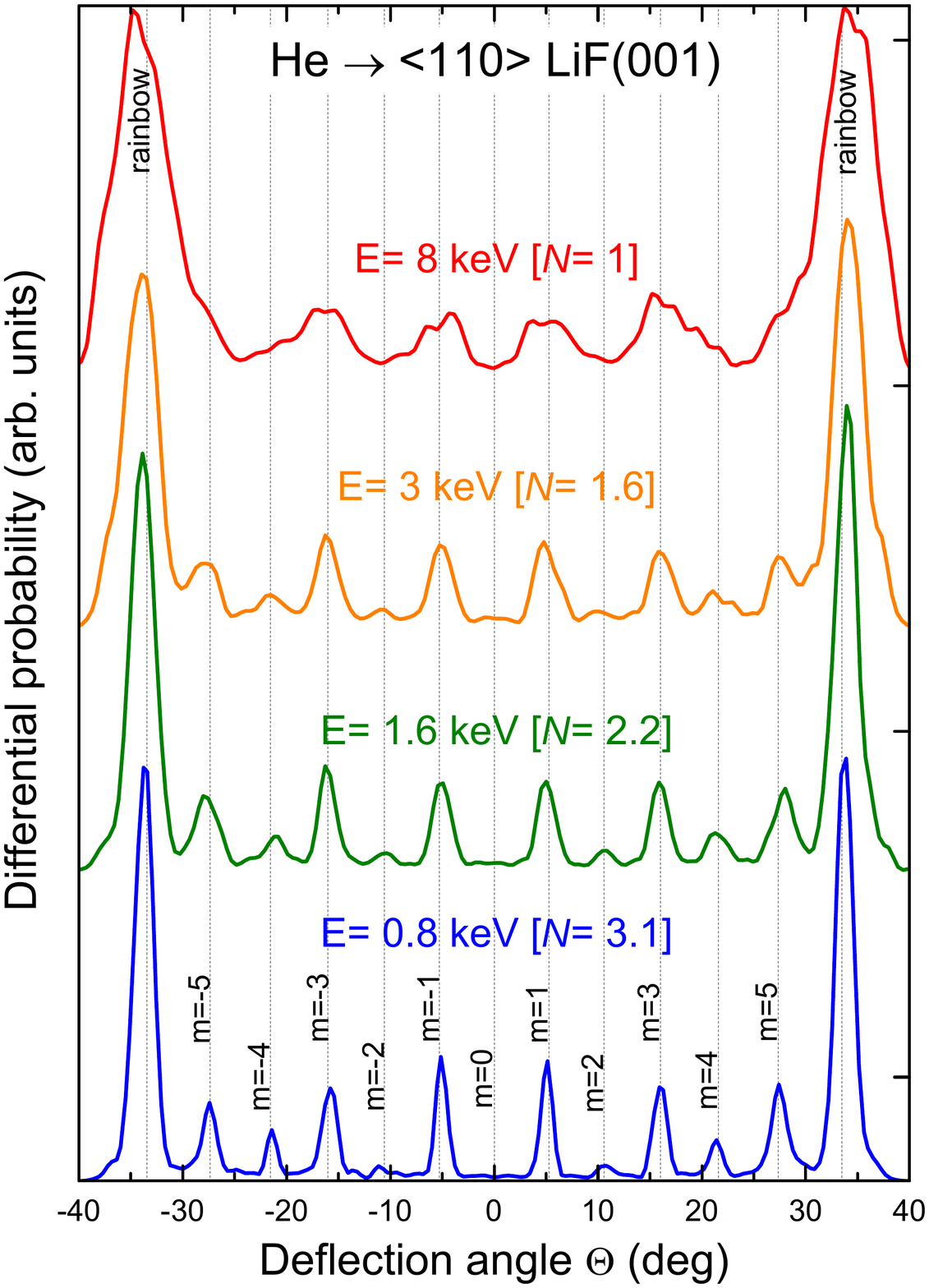}
\caption{(Color online) Angular spectra, as a function of the deflection
angle $\Theta $, for the cases considered in Fig. \protect\ref{hemap} (a).
Dashed vertical lines, $\Theta _{m}$- positions of Bragg peaks.}
\label{hespectra}
\end{figure}

In order to investigate thoroughly the energy dependence of the projectile
distributions displayed in Fig. \ref{hemap} (a), \ in Fig. \ref{hespectra}
we plot the corresponding SIVR differential probabilities as a function of
the deflection angle $\Theta =\arctan (\varphi _{f}\ /\theta _{f})$. Under
ideal scattering conditions, involving the incidence of transversely
extended wave packets,\ these $\Theta $- distributions are expected to be
independent of $E$ at the same $E_{\perp }$ \cite%
{Schuller2009,Gravielle2015b}. Nevertheless, in concordance with Fig. \ref%
{hemap} (a), we remarkably found that the spectra of Fig. \ref{hespectra}
are severely affected by the total energy if the same collimating setup is
used in all the cases. \ For $E=0.8$ keV [$N=3.1$] the projectile
distribution as a function of the deflection angle displays well defined
Bragg peaks, placed at the angular positions $\Theta _{m}$ (indicated with
vertical dashed lines in Fig. \ref{hespectra}) satisfying $\sin \Theta
_{m}=m\lambda _{\perp }/a_{y}$, \ where $m=0,\pm 1,\pm 2,..$ denotes the
Bragg order. But\ since the width of the Bragg peaks depends on $N$ \cite%
{Gravielle2014}, these Bragg structures fade out progressively as the energy
increases, bringing into light supernumerary rainbows, as observed for $E=8$
keV at the top of Fig. \ref{hespectra}. In addition, all the spectra of Fig. %
\ref{hespectra} display high-intensity rainbow maxima at the outermost
angles, which have a classical origin \cite{Guantes2004}.

\subsection{Influence of the projectile mass}

\begin{figure}[tbp]
\includegraphics[width=0.5\textwidth]{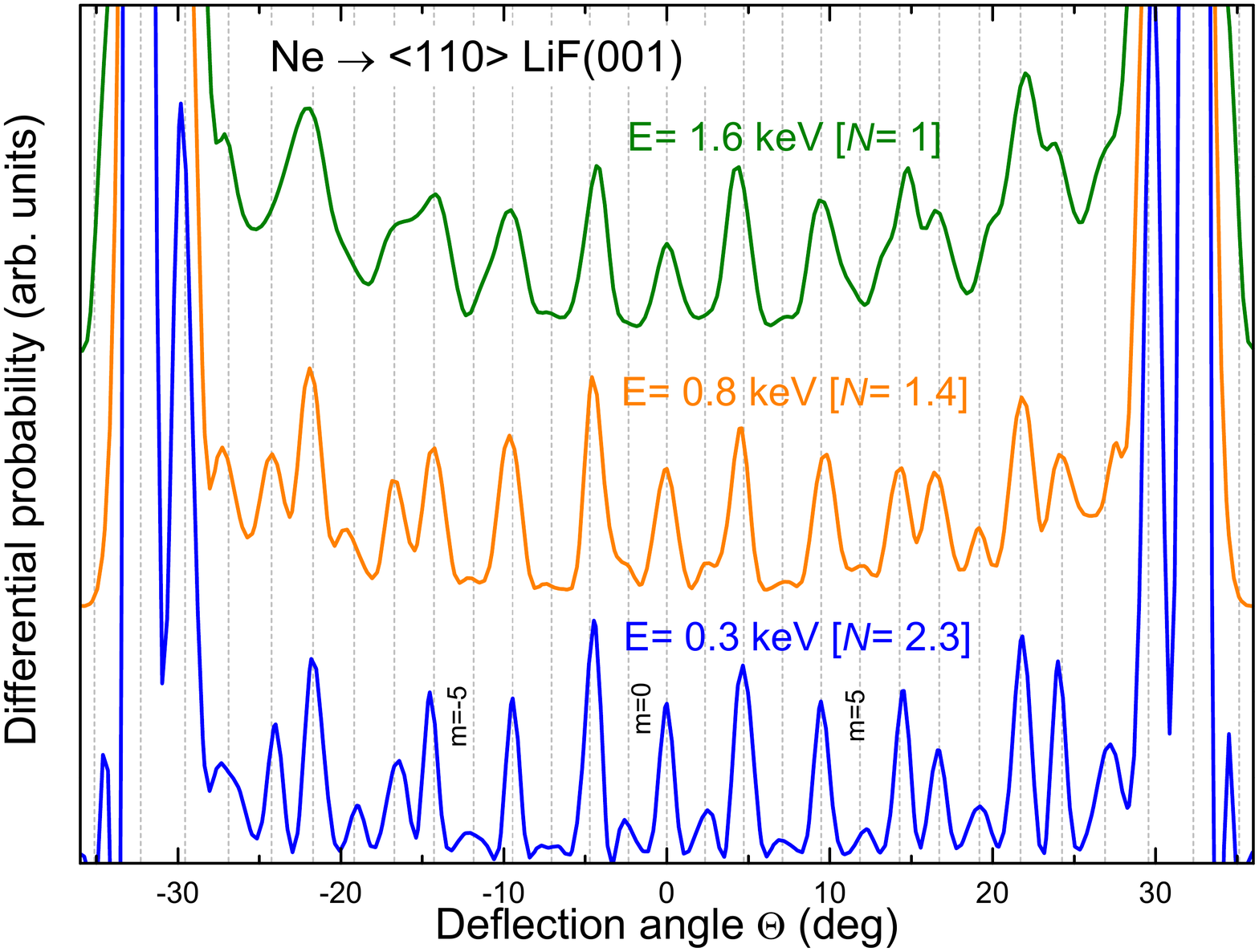} \centering
\caption{(Color online) Analogous to Fig. \protect\ref{hespectra} for Ne
atoms impinging on LiF(001) along the $\left\langle 110\right\rangle $
direction, with $E_{\bot }=0.3$ eV.}
\label{nespectra}
\end{figure}

The angular distributions of neon atoms scattered along the $\left\langle
110\right\rangle $ channel, plotted in Fig. \ref{nespectra}, display a
behavior analogous to that shown in Fig. \ref{hespectra} for helium.
However, for Ne projectiles the dependence of $N$ on the atomic mass,
through the initial momentum as given by Eq. (\ref{ny}), originates a
reduction of the number of coherently lighted channels in comparison with He
at the same impact energy. Therefore, under the same collimating conditions
the limit energy for the observation of inter-channel interference in Ne
spectra results to be about $5$ times lower than in the case of He impact.
Hence, in Fig. \ref{nespectra} the Ne distribution for $E=1.6$ keV shows
only supernumerary rainbow maxima, which contrasts with the Bragg structures
of Fig. \ref{hespectra} for the same impact energy of He projectiles. Notice
that in Fig. \ref{nespectra} well-resolved Bragg peaks are only present in
the Ne distribution for $E=0.3$ keV [$N=2.3$], which is comparable to that
for $1.6$ keV He projectiles in Fig. \ref{hespectra}. These results suggest
that the transverse coherence length might be the central parameter that
limits the observation of Bragg peaks in experimental Ne spectra, rather
than the thermal vibrations of the surface atoms or the spatial resolution
of the detector, as it was previously considered \cite{Gravielle2011}.

\subsection{Contribution of the spot-beam effect}

\begin{figure*}[tbp]
\includegraphics[width=0.9 \textwidth]{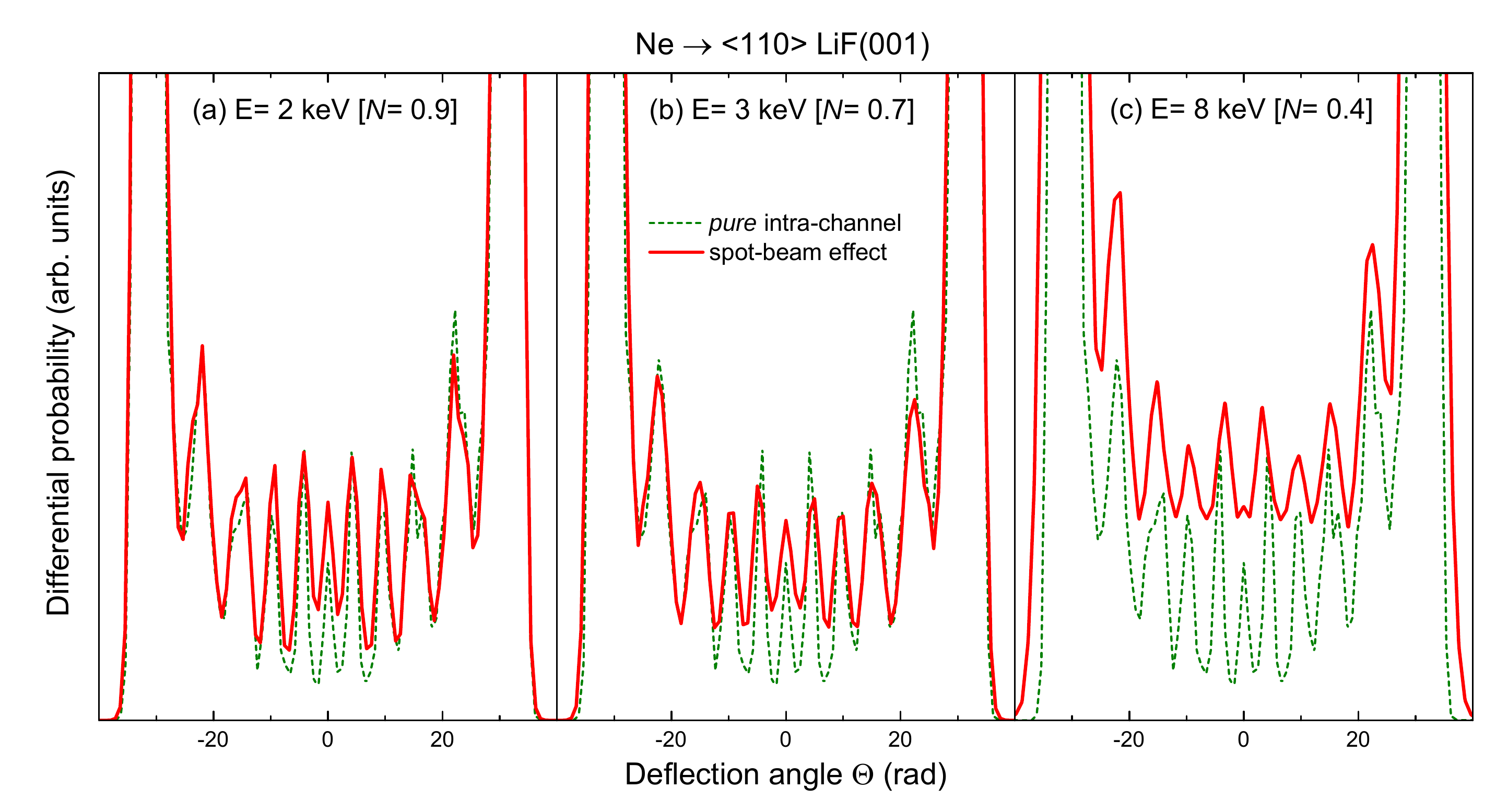}
\caption{(Color online) Angular spectra, as a function of the deflection
angle $\Theta $, for Ne atoms along the $\left\langle 110\right\rangle $
direction, with $E_{\bot }=0.3$ eV. Results for (a) $E=2$ keV $[N=0.9]$, (b)
$E=3$ keV $[N=0.7]$, and (c) $E=8$ keV $[N=0.4]$ are displayed. Red solid
line, angular distribution including the spot-beam effect, as given by Eq. (%
\protect\ref{proba}); dark-green dashed line, \emph{pure} intra-channel
distribution corresponding to $N=1$, given by Eq. (\protect\ref{central}).}
\label{ne-ny09}
\end{figure*}

All the results presented in the previous Sections were obtained from
coherently illuminated regions with a transverse length longer than or equal
to the channel width, that is, with $N\gtrsim 1$. \ Under such a constraint,
the SIVR transition amplitudes corresponding to different focus points of
the beam, given by Eq. (\ref{aif}), are alike, leading to
\begin{equation}
dP^{{\small (SIVR)}}/d\Omega _{f}\simeq \left\vert A_{if}^{{\small (SIVR)}}(%
\mathbf{R}_{s}=0)\right\vert ^{2},  \label{central}
\end{equation}%
where $\mathbf{R}_{s}=0$ indicates a focus point situated just in the middle
of the incidence channel, here named central focus point.

But when the impact energy augments beyond the limit of \textit{pure}
intra-channel interference, and consequently, the coherently lighted area
shrinks, covering a surface region narrower than $a_{y}$, the different $%
Y_{s}$ coordinates of the focus points give rise to dissimilar partial
projectile distributions $\left\vert A_{if}^{{\small (SIVR)}}(\mathbf{R}%
_{s})\right\vert ^{2}$. Each of these partial distributions probes a
different zone of the atom-surface potential within the channel, causing the
contribution of the spot-beam effect, associated with the $\mathbf{R}_{s}$%
-integral in Eq. (\ref{proba}), to become important.

In order to study the energy dependence of the spot-beam effect, in Fig. \ref%
{ne-ny09} we show $dP^{{\small (SIVR)}}/d\Theta $, as a function of the
deflection angle, for Ne atoms impinging along the $\left\langle
110\right\rangle $ channel with total energies (a) $E=2$, (b) $3$, and (c) $%
8 $ keV, which correspond to $N=0.9$ , $0.7$, and $0.4$, respectively. In
all the panels, results derived from Eq. (\ref{proba}), including the
spot-beam contribution, are contrasted with those obtained by considering
only \textit{pure} intra-channel interference, as given by Eq. (\ref{central}%
) for $N=1$. From Fig. \ref{ne-ny09} we found that the spot-beam
contribution keeps the angular positions of supernumerary maxima, but
introduces a non-coherent background in the central region of the spectrum,
around the direction of specular reflection (i.e., $\Theta \simeq 0$), in
relation to that for single-channel illumination. The angular extension of
such a spot-beam background is sensitive to $N$, increasing as $N$
diminishes, as observed by comparing Figs. \ref{ne-ny09} (a) and (b).

The role played by the spot-beam effect is even more relevant when the
transverse length of the surface area that is coherently illuminated by the
beam is about or smaller than the half width of the incidence channel. In
Fig. \ref{ne-ny09} (c) the projectile distribution for $8$ keV Ne atoms
(i.e., $N=0.4$) is severely affected by the spot-beam effect when it is
contrasted with that corresponding to $N=1$. Different $\mathbf{R}_{s}$
positions allow projectiles to separately explore zones of the potential
energy surface with positive or negative slope, \ producing interference
structures placed at negative or positive deflection angles, respectively.
Only when these partial contributions are added, as given by Eq. (\ref{proba}%
), the angular spectrum including the spot-beam contribution presents
defined supernumerary peaks in the whole angular range. But in this case the
spot-beam effect gives also rise to a wide non-coherent background, which
reduces the visibility of the interference patterns, in comparison with that
of the \textit{pure} intra-channel spectrum, as it will be discussed in the
next Section.

\subsection{Focusing effect in the transition from quantum to classical
distributions}

In this Section we investigate how the decreasing of $N$ below the unit
reduces the visibility of the diffraction patterns, leading to the gradual
switch from quantum projectile distributions, containing intra-channel
interference structures, to classical spectra without signatures of
interference. With this aim it is convenient to analyze the profile of the
atom-surface potential near the reflection region of projectiles, which
governs the intra-channel interference in a first approach. Beforehand, we
stress that our SIVR calculations were obtained from a three-dimensional
atom-surface potential and no dimension reduction was made during the
dynamics. However, since FAD patterns are essentially sensitive to the
averaged potential energy surfaces along the incidence channel, such
effective equipotential contours provide useful insights of the
intra-channel interference mechanism.

\begin{figure}[tbp]
\includegraphics[width=0.4\textwidth]{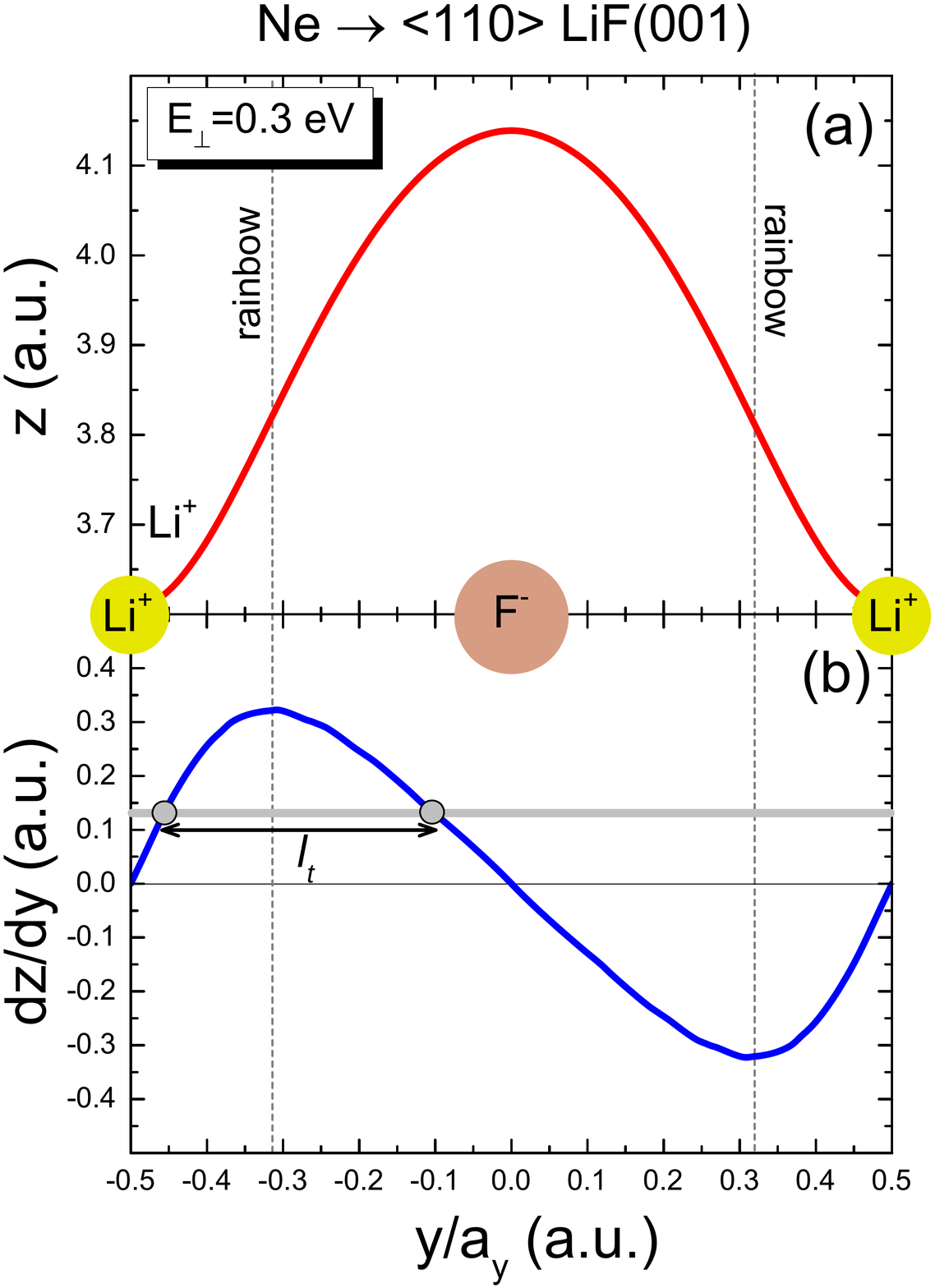}
\caption{(Color online) Analysis of the equipotential contour, averaged
along the $\left\langle 110\right\rangle$ channel, for the Ne-LiF(001)
interaction. (a) Red solid line, equipotential curve $z(y)$ for $E_{\bot
}=0.3$ eV; (b) derivative $dz/dy$ of the equipotential curve of (a). Gray
circles, turning point positions corresponding to two different trajectories
that interfere at a given deflection angle; vertical dashed lines indicate
turning point positions corresponding to trajectories that contribute to the
rainbow maxima.}
\label{equip}
\end{figure}

\begin{figure}[tbp]
\includegraphics[width=0.5\textwidth]{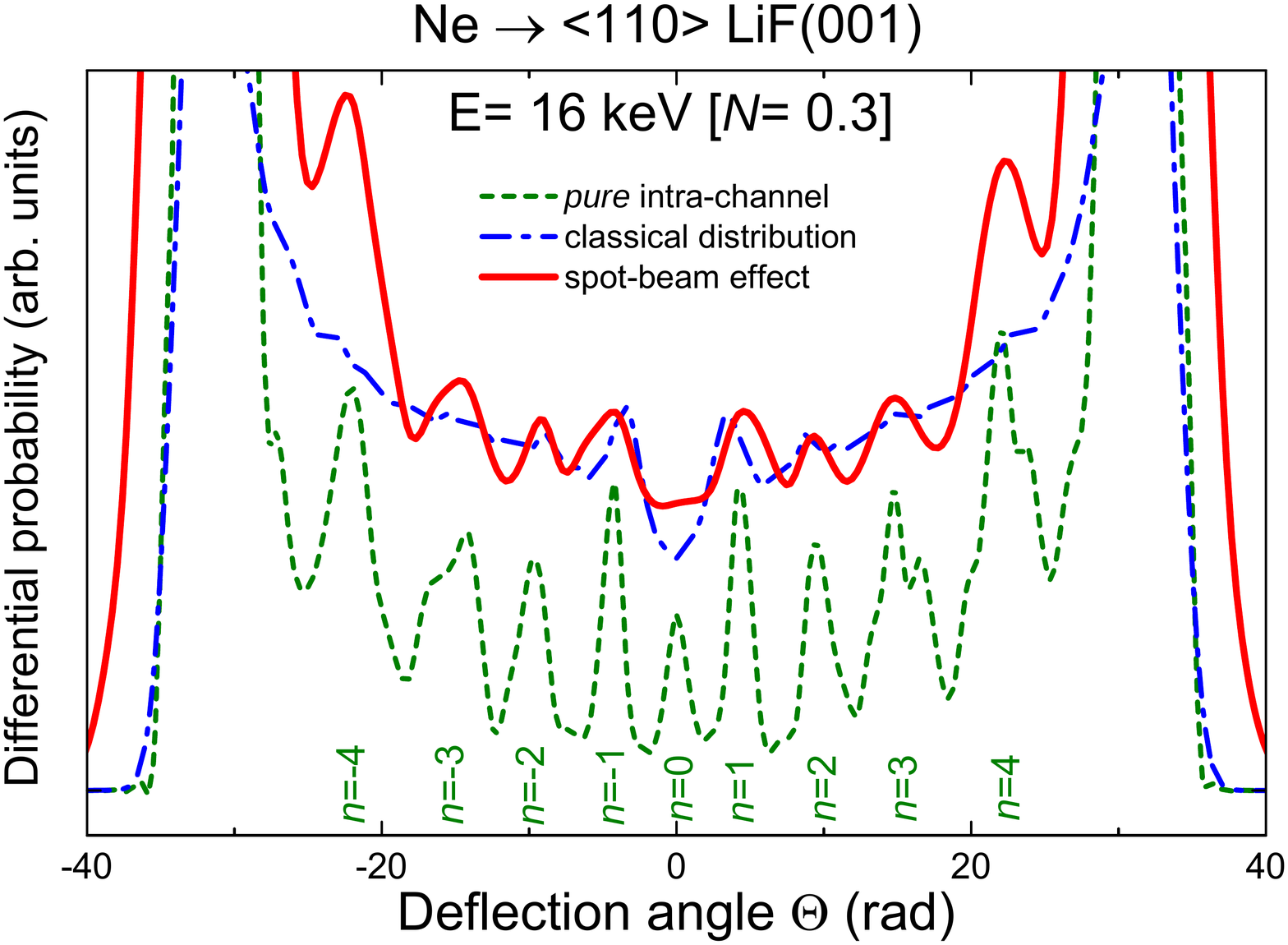}
\caption{(Color online) Analogous to Fig. \protect\ref{ne-ny09} for $E=16$
keV $[N=0.3]$. Blue dot-dashed line, classical projectile distribution for$%
N=1$. The $n$ values indicate different supernumerary rainbow peaks.}
\label{ne-ny03}
\end{figure}

For Ne atoms impinging on LiF(001)\ along the $\left\langle 110\right\rangle
$ direction, in Fig. \ref{equip} (a) we plot the averaged equipotential
curve - $z(y)$ - corresponding to $E_{\perp }=0.3$ eV, as a function of the
coordinate $y$ across the channel, normalized by the width $a_{y}$. Within
this simplified picture, the intra-channel interference is produced by the
coherent\ addition of transition amplitudes $a_{if}^{{\small (SIVR)}}(%
\mathbf{r}_{o},\mathbf{k}_{o})$ corresponding to trajectories reflecting at
turning points with different $y$ coordinates inside the channel, but with
the same slope $dz/dy$ of the averaged equipotential curve, which determines
the final azimuthal angle \cite{Winter2011}. For the present case, from Fig. %
\ref{equip} (b) it is observed that there are only two different
trajectories that contribute to the intra-channel interference pattern at a
given angular position $\varphi _{f}$, except around rainbow angles where
several (infinite) turning points coalesce at a maximum or minimum of $dz/dy$%
. Then, in FAD distributions an essential requirement to observe a
supernumerary rainbow structure at a given $\varphi _{f}$ or $\Theta $ angle
(inside the angular range defined by the rainbow peaks) is given by the
condition that the $N$ value must be longer than $l_{\mathrm{t}}=l_{y}/a_{y}$%
, where $l_{y}$ denotes the transverse distance between the turning points
of the corresponding interfering trajectories. This fact is illustrated in
Fig. \ref{ne-ny03}, where the distribution for $\left\langle
110\right\rangle $ incidence of $16$ keV Ne atoms, corresponding to $N=0.3$,
is displayed. In this case the spectrum obtained including the spot-beam
effect does not show the central interference maximum, associated with the
highest $l_{\mathrm{t}}$ value (i.e., $l_{\mathrm{t}}=0.5$ corresponding to
two trajectories on top of the Li$^{+}$- and F$^{-}$- crystallographic rows,
respectively), and supernumerary rainbow peaks are visible at larger
deflection angles only. \ In addition, the visibility of the interference
structures diminishes and the spectrum tends to the classical distribution,
which displays only pronounced rainbow maxima.

\begin{figure}[tbp]
\includegraphics[width=0.5\textwidth]{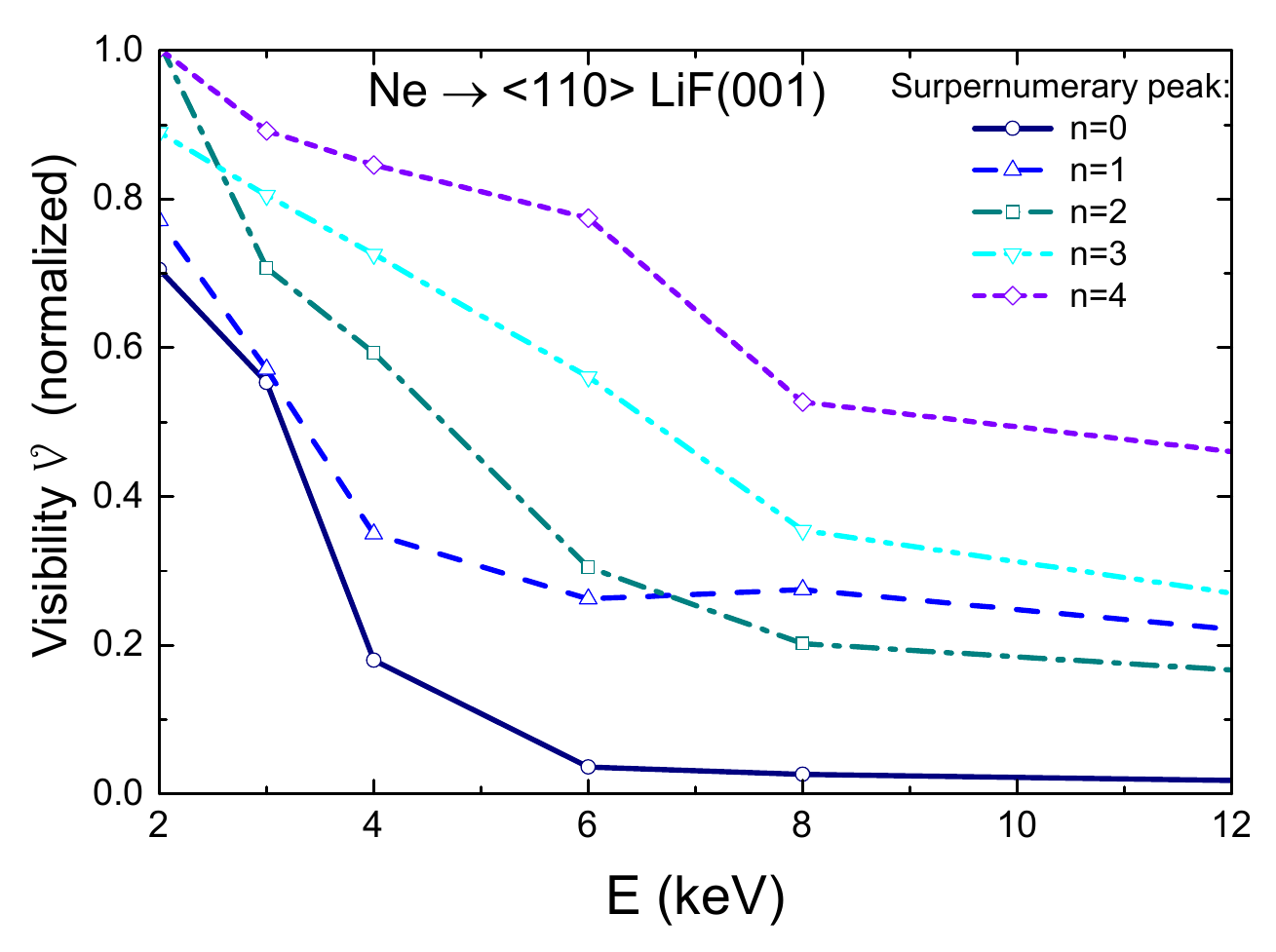}
\caption{(Color online) Visibility $\mathcal{V}(n)$ (normalized to that for $%
N= 1$), as a function of $E$, for Ne projectiles colliding along $%
\left\langle 110\right\rangle $ with $E_{\perp }=0.3$ eV.}
\label{visibE}
\end{figure}

\begin{figure*}[tbp]
\includegraphics[width=0.8 \textwidth]{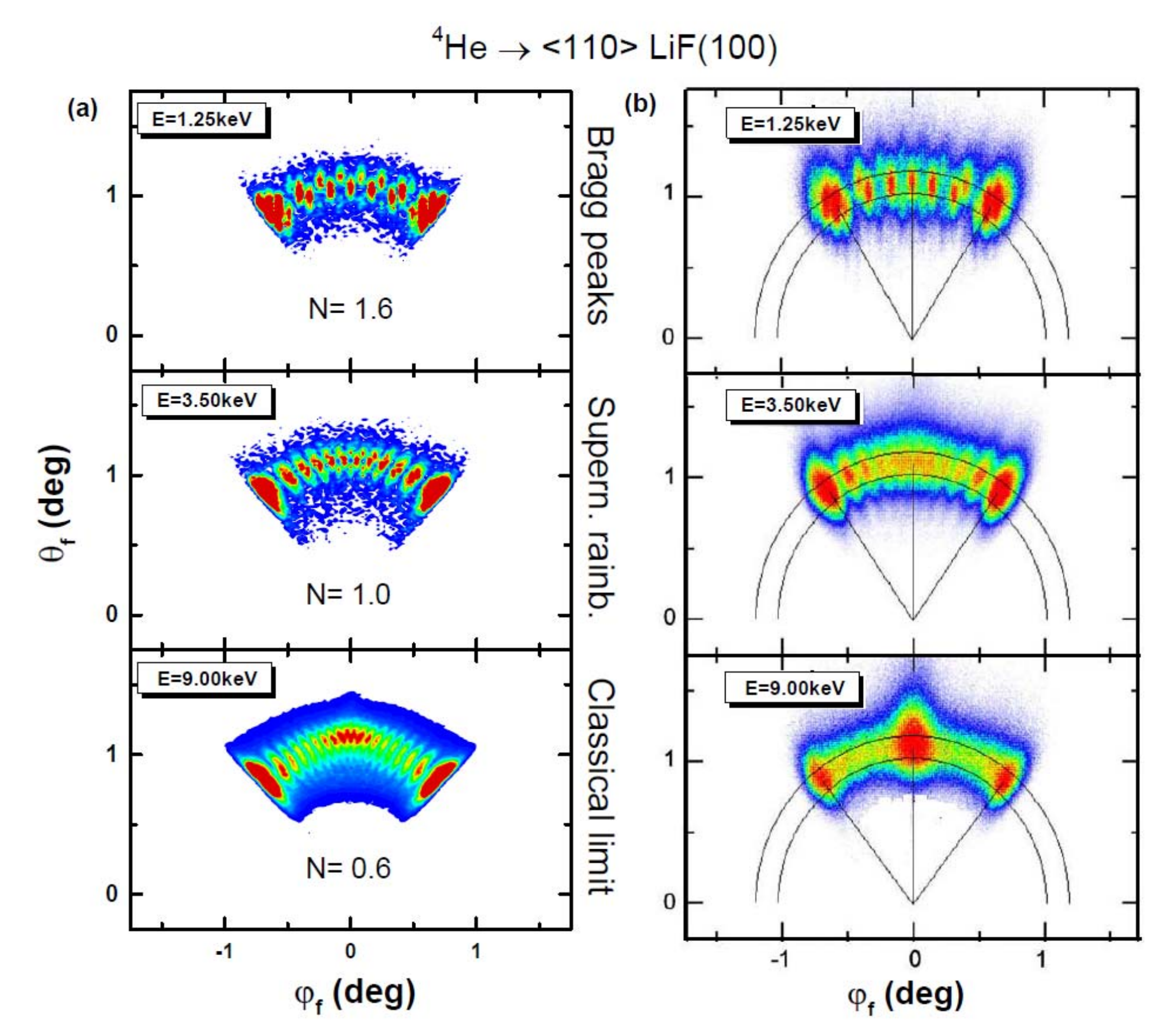}
\caption{(Color online) Two-dimensional angular distributions, as a function
of $\protect\theta _{f}$ and $\protect\varphi _{f}$, for He atoms impinging
on LiF(001) along $\left\langle 110\right\rangle $ with $\protect\theta %
_{i}=1.1\deg $. (a) (right panels) Simulated SIVR results derived by
considering a collimating slit of width $d_{y}=0.3$ mm. (b) (left panels)
Experimental distributions extracted from Ref. \protect\cite{Schuller2009c}.
In both columns, different impact energies - $E=$ $1.25$, $3.50$, and $9.00$
keV - are considered. The corresponding $N$ values, as given by Eq. (\protect
\ref{ny}), are indicated.}
\label{expt}
\end{figure*}

The quantum-classical transition of FAD distributions can be quantitatively
studied by analyzing the visibility $\mathcal{V}(n)$ associated with the
supernumerary rainbow maximum labelled with $n$ in Fig. \ref{ne-ny03}, where
$n=0,\pm 1,\pm 2,..$, $n=0$ corresponding to the central peak \cite{note}.
Like in optics \cite{BornWolf}, we define the visibility in FAD as

\begin{equation}
\mathcal{V}(n)=\frac{I_{\max }^{(n)}-I_{\min }^{(n)}}{I_{\max
}^{(n)}+I_{\min }^{(n)}},  \label{visib}
\end{equation}%
where $I_{\max }^{(n)}$ is the differential probability $dP^{{\small (SIVR)}%
}/d\Theta $, derived from Eq. (\ref{proba}), at the $n$- supernumerary
rainbow maximum, and $I_{\min }^{(n)}$ denotes the averaged value of the
differential probability at the positions of the two adjacent minima. This
visibility provides a measure of the degree of coherence of the atomic beam
\cite{BornWolf,Adams1994}. In Fig. \ref{visibE} we show $\mathcal{V}(n)$,
normalized to that for $N=1$, as a function of the impact energy, for Ne
projectiles colliding along $\left\langle 110\right\rangle $ with $E_{\perp
}=0.3$ eV. As a consequence of the spot-beam effect, under the same
collimating conditions the visibility tends to decrease when the energy
increases beyond the energy limit of \textit{pure} intra-channel
interference. Such a decreasing is more steeply for the central peaks than
for the outermost ones, in accord with the condition $N\gtrsim l_{\mathrm{t}%
} $ for the observation of supernumerary maxima. For higher $E$ (lower $N$)
values, the interference structures gradually blur out and all $\mathcal{V}%
(n)$ slowly decrease, making projectile distributions reach the classical
limit.

\subsection{Experimental comparison}

To test the predicted influence of the incidence conditions, in Fig. \ref%
{expt} SIVR simulations derived from Eq. (\ref{proba}) are compared with
available experimental distributions \cite{Schuller2009c} for helium atoms
impinging on LiF(001) along the $\left\langle 110\right\rangle $ channel.
These two-dimensional angular distributions were obtained by varying the
impact energy but keeping fixed the incidence angle, i.e., $\theta
_{i}=1.1\deg $ \cite{Schuller2009c}. In order to reproduce the experiments,
in this Section we have considered a rectangular slit with sides $d_{x}=1.5$
mm and $d_{y}=0.3$ mm, which produces an angular dispersion $\omega
_{\varphi }=0.05\deg $, comparable to the experimental value \cite%
{Schuller2009c}.

In Fig. \ref{expt}, for $E=1.25$ keV (top panels) the theoretical
distribution is in accord with the experimental one, showing not fully
resolved Bragg peaks associated with $N=1.6$. \ Instead, for $E=3.50$ keV
(middle panels) the Bragg peaks completely disappear and the simulated and
experimental FAD patterns display only supernumerary maxima corresponding to
a single-channel illumination, i.e., to $N=1$. \ Lastly, for $E=9.00$ keV [$%
N=0.6$] (lower panels) the interference maxima are barely visible as
isolated peaks in the simulated angular spectrum due to the contribution of
the spot-beam effect. In this case, both the theoretical and the
experimental distributions tend to the classical one, showing a broad high
intensity contribution at $\varphi _{f}=0$ and two intense rainbow peaks at
the outermost angles.

Therefore, the reasonable good agreement between theory and experiment
observed in Fig. \ref{expt} strongly suggests that the energy dependence of
the general features of experimental FAD distributions is mainly produced by
the variation of the transverse coherence length, as it was proposed in Ref.
\cite{Schuller2009c}. However, at this point it is necessary to mention that
there are other effects not included in our model, like inelastic processes
\cite{Roncin2017}, which can contribute to deteriorate the coherence,
helping to the transition from quantum to classical projectile
distributions. In addition, notice that our simulations do not include
thermal vibrations of lattice atoms \cite{Miraglia2017} and the results were
not convoluted with the detector resolution, both effects which are expected
to smooth the theoretical spectra.

\section{Conclusions}

We have investigated the influence of the total energy, the incidence
channel, and the projectile mass on the general characteristics of FAD
patterns produced by an atomic beam that collides grazingly on a LiF(001)
surface, after passing through a fixed collimating setup. We have shown
that, even using the same collimating aperture, it is possible to obtain
final projectile distributions containing\ different interference structures
by varying the total energy while keeping the normal energy as a constant.\
This behavior can be explained in terms of the number $N$ of equivalent
parallel channels that are coherently illuminated by the atomic beam.

For a given collimating aperture, the decreasing of $N$ as $E$ increases,
originates the gradually broadening of Bragg peaks, until they fade
completely out for $N=1$, bringing to light supernumerary rainbow maxima
associated with the \textit{pure} intra-channel interference. The energy
ranges corresponding to these interference mechanisms also depend on the
projectile mass, making \textit{pure} intra-channel interference be reached
at a total energy lower for Ne than for He projectiles, as it was
experimentally observed \cite{Gravielle2011}.

Furthermore, we have demonstrated that the spot-beam effect, produced by the
different positions within the crystal lattice of the focus point of the
beam, plays an important role when a portion of a single crystallographic
channel is coherently lighted by the impinging particles. In the energy
range where $0.4\lesssim $ $N$ $\lesssim 1$ the spot-beam effect helps to
recover supernumerary rainbow maxima symmetrically distributed with respect
to the specular direction. But the visibility of these structure decreases
as $N$ diminishes, causing that for smaller $N$ values the projectile
distributions approximate to the classical ones.

In conclusion, the focusing effects have be shown to be essential to
properly describe \ experimental projectile distributions, which is relevant
to use FAD spectra as a surface analysis tool. In this regard, notice that
the coherence conditions of the incident beam have been recently found to
play an important role in \ atomic collisions involving not only molecules
\cite{Egodapitiya2011} but also atoms \cite{Gassert2016,Sarkadi2016} as
targets. Hence, FAD distributions might represent an almost ideal benchmark
to investigate this problem, being a useful guide for coherence studies on
other collision systems. Finally, we should mention that even though present
results are in agreement with available experiments \cite{Schuller2009c},
further experimental research on this issue is desirable.

\begin{acknowledgments}
This work was carried out with financial support from CONICET, UBA and
ANPCyT of Argentina.
\end{acknowledgments}


\end{document}